\documentclass[draft,numberedheadings]{aipproc}

\layoutstyle{6x9}

\begin{document}

\def\simg{\mathrel{\rlap{\raise 0.511ex \hbox{$>$}}{\lower 0.511ex \hbox{$\sim$}}}}
\def\siml{\mathrel{ \rlap{\raise 0.511ex \hbox{$<$}}{\lower 0.511ex \hbox{$\sim$}}}}

\title{Direct and bulk-scattered forward-shock emissions: sources of X-ray afterglow diversity}

\classification{ }
\keywords{gamma-ray bursts, shock-waves, non-thermal emission}

\author{A. Panaitescu}{ 
   address={ ISR-1, Los Alamos National Laboratory, Los Alamos, NM 87545, USA} }

\begin{abstract}
 I describe the modifications to the standard forward-shock model required to account 
for the X-ray light-curve features discovered by Swift in the early afterglow emission
and propose that a delayed, pair-enriched, and highly relativistic outflow, which
bulk-scatters the forward-shock synchrotron emission, yields sometimes a brighter
X-ray emission, producing short-lived X-ray flares, X-ray light-curve plateaus ending 
with chromatic breaks, and fast post-plateau X-ray decays.
\end{abstract}

\maketitle


 The hundreds of X-ray afterglow light-curves measured by Swift/XRT during the last years 
display three phases of power-law decay $F_x \propto t^{-\alpha_x}$: a plateau (slow-decay), 
a normal decay, and a steep fall-off (e.g. \cite{zhang06}). Only rarely, the 0.3-10 keV flux
exhibits an extremely rapid drop by 1-2 decades at the end of the plateau.

{\bf Normal decay}. During the second phase, the X-ray decay index $\alpha_x \in (0.75,1.5)$ 
is similar to that measured for $\sim 40$ pre-Swift optical light-curves \cite{zeh06},  
a majority (70-90\%) of X-ray decay indices being compatible \cite{alin1} with the 
expectations of the {\sl standard} {\sc forward-shock} model (e.g. \cite{meszaros97}), 
for the measured X-ray spectral slope $\beta_x$ (with $F_\nu \propto \nu^{-\beta_x}$). 
In this model, the afterglow is identified with the synchrotron emission from ambient 
electrons accelerated to energies $\gamma m_e c^2 \sim$ GeV energies (comoving frame) 
at the ultrarelativistic shock driven by the GRB ejecta into the circumburst medium.

 In its standard form, the forward-shock model assumes that (1) no energy is added to the 
blast-wave, (2) its kinetic energy is uniformly distributed with angle ($dE_k/d\Omega=0$), 
and (3) electrons and magnetic fields acquire fractions ($\varepsilon_e$ and $\varepsilon_B$) 
of the post-shock energy that are constant.
With these assumptions, the power-law deceleration of the blast-wave (Lorentz factor
$\Gamma \propto t^{-g}$, where $g$ depends on the ambient medium radial stratification)
and the power-law distribution of shock-accelerated particles ($dN/d\gamma \propto
\gamma^{-p}$) lead to a power-law decay of the
synchrotron flux at photon energies above the peak of the synchrotron spectrum, with
$\alpha_x = 1.5 \beta_x + c$. Compatibility of the $\alpha_x$ and $\beta_x$ measured
for Swift X-ray afterglows and the standard forward-shock model predictions is obtained 
if the X-ray domain is, for some afterglows, below the cooling frequency of the synchrotron 
spectrum (see \cite{sari98} for definition) and above it for others. 

{\bf Steep fall-off}. For this phase, $\alpha_x \in (1.75,2.75)$, which is compatible 
with the post-break decay indices measured for $\sim 15$ pre-Swift optical light-curves 
\cite{zeh06}. Such a light-curve break was predicted to arise from the finite angular 
extent of the forward-shock \cite{rhoads99} and was identified in about 3/4 of pre-Swift 
optical afterglows well-monitored until after a few days.
 There has been a suggestion \cite{burrows07} that Swift X-ray afterglows do not display 
"jet-breaks" as often as pre-Swift optical light-curves. In a sample of about 100 Swift 
X-ray afterglows with good temporal coverage, I find roughly equal numbers of light-curves 
with good evidence for a jet-break and without a break until at least a few days, 
indicating that the fraction of X-ray light-curves with jet-breaks is around 1/2 \cite{alin2}. 
The discrepancy between the fractions of X-ray and optical light-curves with jet-breaks
may be due to the jet-breaks of Swift X-ray afterglows occurring somewhat later 
(0.3-10 days) than for pre-Swift optical afterglows (0.3-3 days) and, thus, requiring 
a longer monitoring to catch the break. 

 Alternatively, that jet-breaks occur less frequent in Swift X-ray afterglows may be 
an indication of a departure from the standard forward-shock model and that, perhaps,
the optical and X-ray emissions have different origins (as discussed below). 

 Before that, one should check first if the standard forward-shock model can account 
for the X-ray light-curve pre- and post-break decay indices and if the X-ray breaks 
are {\sl achromatic} (i.e. if they occur in the optical as well). Sadly, such tests 
are not very conclusive: for those X-ray light-curves with potential jet-breaks, 
the pre- and post-break decay indices cannot be reconciled with the spectral slope 
within a single variant of the standard forward-shock model, but most (90\%) can be 
accommodated if, e.g., some jets expand laterally while others do not \cite{alin2}; 
simultaneous optical and X-ray coverage after 1 day has been acquired for only a few 
afterglows with potential X-ray light-curve "jet-breaks", the overall temporal coverage 
being, in general, insufficient for a robust test of break achromaticity.  

 {\bf Plateau}. A stronger evidence that the standard forward-shock model does not 
provide a complete description of the afterglow emission comes from the plateau phase,
as described below.

 Relaxing any of the above three assumptions of the standard forward-shock model can 
explain the slow-decay phase. For the first two, an increasing kinetic energy per solid 
angle over the ever-increasing area (of angular extent $\theta = \Gamma^{-1}$) "visible" 
to the observer can be acquired either by energy being injected into the forward shock 
(by means of some initially-slower ejecta catching-up later with the decelerating shock) 
or by a shock region of higher $dE_k/d\Omega$ becoming visible to the observer 
("structured outflow").

 The structured outflow scenario \cite{eichler06} seems disfavoured by that the correlations 
expected among the flux, emergence epoch, and decay index are not manifested by Swift 
X-ray plateaus \cite{alin1}. 

 The energy injection scenario encounters a difficulty in explaining the decoupling
of the optical and X-ray light-curves observed for some Swift afterglows (Figure 1),
as the cessation of energy injection in the shock should lead to comparable increments 
$\delta \alpha$ of the optical and X-ray decay indices: analytically, one expects that
$\delta \alpha_x - \delta \alpha_o \in (-0.75,0.25)$. In particular, {\sl chromatic}
X-ray breaks (i.e. which are not manifested in the optical) are most troubling for this
scenario because cessation of energy injection (at the end of the X-ray slow decay phase) 
alters the forward-shock dynamics and should show an effect at all wavelengths.
\footnote{
That the optical light-curve does not display a break at the same time would suggest 
that the X-ray break is caused by the cooling frequency crossing the X-ray domain 
(provided that it can evolve so fast as to yield the observed X-ray steepening 
$\delta \alpha_x$), however X-ray plateau ends are not accompanied by a spectral 
evolution \cite{nousek06} \cite{willingale07}}

\begin{figure}
  \includegraphics[height=.3\textheight]{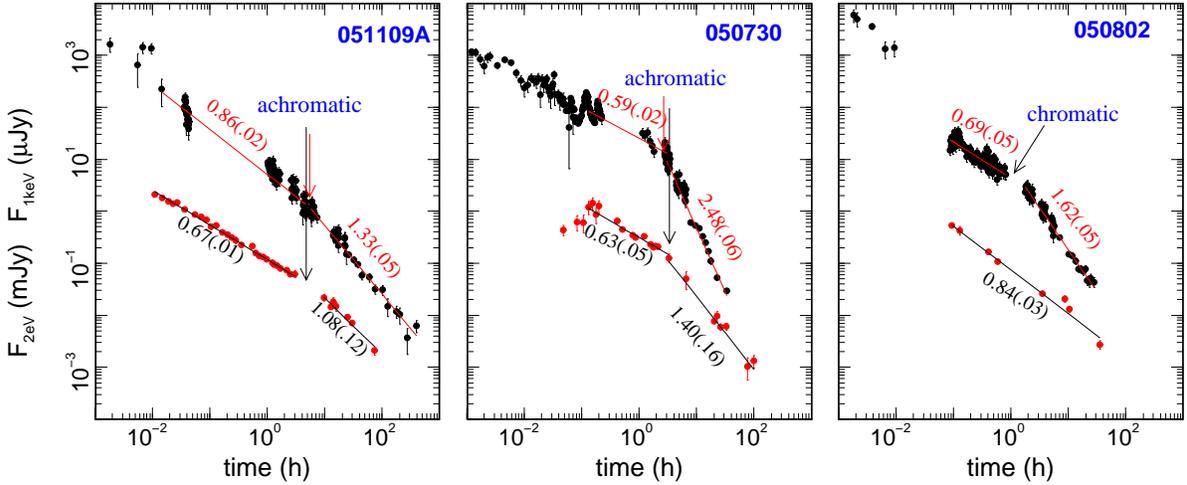}
  \caption{ Various degrees of coupling manifested by optical and X-ray afterglow 
   light-curves: left panel shows an achromatic break for which the pre- and post-break
   optical and X-ray decay indices differ by $\siml 1/4$ (which is within the reach of
   the standard forward-shock model), mid panel shows an achromatic break for which 
   the post-break decay indices differ substantially, right panel shows a chromatic 
   X-ray break which is not seen in the optical (the breaks of the last two cases cannot
   be explained with only cessation of energy injection). If the slow decay phase
   before the break is identified with energy injection in the forward shock and the
   break with cessation of that process then the increasing degree (from left to right) 
   of decoupling of optical and X-ray light-curves requires a stronger evolution of 
   microphysical parameters.
   Numbers indicate the flux power-law decay index $\alpha$ and its $1\sigma$ uncertainty.  }
\end{figure}

 That X-ray light-curve breaks are stronger than in the optical (in the extreme case
even lacking in the optical) indicates that the remaining assumption of microphysical 
parameters $\varepsilon_e$ and $\varepsilon_B$ constancy must also be relaxed. Then,
the least contrived forward-shock scenario for X-ray plateaus and their ends that can 
be construed is that where energy injection ceases to be dynamically important at the 
plateau end epoch and microphysical parameters have a steady evolution with the Lorentz 
factor across the break. This model has three degrees of freedom (for energy injection
law and evolution of microphysical parameters) and four observational constraints (the
pre- and post-break optical and X-ray decay indices). Interestingly, this overconstrained
model can account for the four observables for a majority (80\%) of the afterglows
analyzed (a smaller set of 6 afterglows is discussed in \cite{alin3}) if the circumburst 
medium has a wind-like stratification, as it should if the progenitors of long GRBs are 
massive stars.
However, there are two puzzling/contrived properties of this model. One is that the 
resulting evolutions of the microphysical parameters lack universality. The second is
that chromatic X-ray breaks are seen at the end of plateau quite often, in about 1/2 of 
the $\sim 15$ afterglows that were also followed in the optical \cite{liang08}, which 
requires that the evolutions of microphysical parameters are often correlated in a way that 
"irons out" the optical light-curve break caused by the changing dynamics of the blast-wave 
when energy injection stops. On the positive side, there is a hint of universality of that
correlation: $(d\ln\varepsilon_e/d\ln\Gamma) + 0.75 (d\ln\varepsilon_B/d\ln\Gamma) = 
-2.6 \pm 0.8$.

{\bf Forward-shock model}.
 To summarize, the most attractive feature of the standard forward-shock model is that
it can explain naturally the power-law decaying afterglow fluxes without further 
assumptions, based only on the blast-wave power-law deceleration and the power-law energy 
spectrum of the radiating electrons. Allowing for injection of energy, this model can 
explain the X-ray plateaus and the achromatic light-curve breaks for which 
$\delta \alpha_x \simeq \delta \alpha_o$. 
The forward-shock model can also accommodate (1) achromatic breaks with different 
$\delta \alpha_o$ and $\delta \alpha_x$ and (2) chromatic breaks provided that microphysical
parameters evolve. However, the latter type of breaks occurs too often, making this model 
too contrived. 

{\bf Other models}. 
 If energy injection into the blast-wave persists for long times, then the 
{\sc reverse-shock} energizing the incoming ejecta can also be long-lived and provide 
an alternate origin for the afterglow emission. In this model, the afterglow light-curve 
reflects the distribution of mass and Lorentz factor in the incoming ejecta. The reverse-shock 
has been shown to produce both coupled and decoupled optical and X-ray light-curves 
\cite{genet07}, although at the cost of a rather contrived feature: a small fraction 
(1\%) of the ejecta electrons acquire a large fraction (50\%) of the dissipated energy. 
When all ejecta electrons are accelerated, chromatic X-ray breaks 
can be obtained if the cooling frequency is between optical and X-ray, the decoupling 
of the light-curves being a consequence of the continuous injection of fresh ejecta 
electrons and their cooling \cite{uhm07}. 

 Although a very rare occurrence, the sharp drops with $\alpha_x \in (3,10)$ seen at 
the end of a couple of X-ray plateaus pose a serious problem to both the reverse and 
forward-shock models because, for an afterglow source radius $R \simeq \Gamma^2 ct$, 
the geometrical curvature of the emitting surface introduces a delay in the arrival-time 
of photons such that the sharpest decay of the blast-wave light-curve is that of the 
"large-angle emission" arriving from angles $\theta > \Gamma^{-1}$, for which 
$\alpha_x = \beta_x + 2 \in (2.5,3.5)$ \cite{kumar00}.

 The short-time fluctuations seen in many X-ray afterglows and the above-mentioned X-ray 
light-curve sharp drops bear similarity with the GRB variability and tail, respectively,
which may suggest that the same mechanism ({\sc internal shocks} ?) operates both 
during the prompt emission phase and the afterglow \cite{ghisellini07}. In this model, 
the "central engine" produces a fluctuating relativistic outflow which radiates at radius 
that must be much smaller than that of the forward shock, to account for the short X-ray 
flares and sharp post-plateau decays. Because the afterglow light-curves depend on the
variation in time of the mass and Lorentz factor of the ejected outflow, it is expected
that the optical and X-ray light-curves from internal shocks are well coupled, thus this
model appears unable to account for chromatic X-ray breaks.

 The different behaviours occasionally displayed by the optical and X-ray light-curves 
may also suggest that, in some cases, the afterglow emissions at these two frequencies 
have different origins. The next natural step is to speculate that the X-ray afterglow 
arises from scattering of the burst and/or afterglow emissions. To account for the much 
longer duration of the afterglow, scattering of the burst emission must introduce a time 
delay, as for {\sc dust-scattering} in the host galaxy \cite{shao07}. Owing to the
finite radial extent of the dust screen, the dust-scattered model has the ability to 
produce X-ray plateaus, however this model also predicts a substantial softening of the
X-ray spectrum ($\delta \beta_x = 3.5$) from plateau to the steep-decay phase and a 
strong dependence of the plateau duration on observing frequency ($t_{plateau} \propto 
E^{-2}$), both of which are in clear contradiction with Swift observations \cite{shen08}.

 Scattering of the afterglow emission could be local or not. Local, {\sc inverse-Compton 
scattering} in the forward-shock has the same abilities/limitations in accounting for
the decoupled optical and X-ray light-curves as the synchrotron emission model, and
requires both energy injection and evolving microphysical parameters \cite{alin3}.

 {\bf Bulk-scattering model}. 
Scattering of the forward-shock emission by a relativistic outflow located behind it 
may overshine the direct synchrotron forward-shock emission in the X-rays but not in 
the optical, thus decoupling the X-ray and optical afterglows \cite{alin4} (Figure 2). 
For this situation to occur, the Lorentz factor of the scattering outflow should be 
substantially larger than that of the forward shock (which can be just a natural 
consequence of that shock's deceleration) and the scattering outflow should either be 
pair-enriched, to acquire a sufficiently high (though below unity) optical thickness 
to electron scattering, or be energized so that its electrons also inverse-Compton 
scatter the photons left behind by the forward-shock.

\begin{figure}
  \includegraphics[height=.27\textheight]{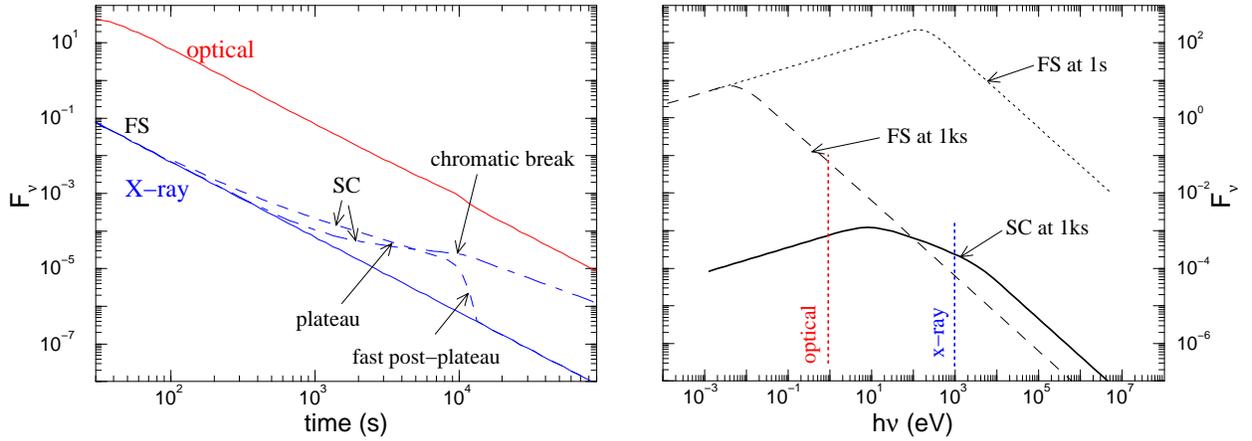}
  \caption{ Left panel: light-curves of forward-shock (FS) and scattered (SC) emissions
     showing that the scattering outflow model can produce plateaus, chromatic X-ray breaks 
     and fast post-plateau decays. Right panel: spectra of forward-shock and scattered
     emissions showing that the scattered emission can overshine the forward-shock's
     at higher photon energies (X-ray).  }
\end{figure}

  Because of the higher Lorentz factor of the scattering outflow, the swept-up 
forward-shock emission arrives at observer on a timescale much shorter than the
observer-frame afterglow age, thus the scattered flux received at time $t$ depends
on the density and Lorentz factor of the scattering outflow at distance $ct$ behind 
the forward-shock at the onset of its deceleration. Consequently, in the scattering
outflow model, the bright and short flares observed in many X-ray afterglows at early
times arise from sheets of higher density and/or higher Lorentz factor in the scattering 
outflow, plateaus are associated with the radially-extended part of the outflow,
and X-ray light-curve breaks are identified with changes in the radial distribution of
the scattering outflow's mass or Lorentz factor.

 While this model can account for many of the novel X-ray light-curve features discovered
by Swift, it has difficulty in explaining achromatic breaks, which would require that 
the optical emission is also scattering-dominated. Instead, achromatic plateau ends are 
more naturally attributed to cessation of energy injection in the forward shock. For these
reasons, we suggest that the degree at which optical and X-ray afterglow light-curves 
are correlated results from the interplay between the bulk-scattered and direct forward-shock 
emissions: chromatic X-ray plateau ends are seen when the former is dominant while achromatic 
plateau ends with $\delta \alpha_x \simeq \delta \alpha_o$ occur when the latter is brighter. 
In this model, achromatic breaks with very different steepenings $\delta \alpha_x$ and 
$\delta \alpha_o$ (of which there is only one -- GRB 050730) should be attributed to the 
forward-shock emission with non-constant microphysical parameters. 

 There are two immediate tests for the above {\sc hybrid} {\sl forward-shock} model. 
First, if only one X-ray emission mechanism is dominant at all times, then {\sl chromatic 
and achromatic breaks should not occur in same afterglow}. So far, that seems to be the case. 
Second, because a chromatic X-ray break occurs only if the scattered emission is brighter 
than the forward shock's, {\sl the X-ray to optical flux ratio $F_x/F_o$ should be, 
on average, larger for afterglows with chromatic X-ray breaks than for those with 
achromatic breaks}. For a set of 15 GRB afterglows with such breaks (7 chromatic, 
7 achromatic), we find that test to be satisfied, the average ratio $F_x/F_o$ at plateau 
end of afterglows with chromatic X-ray breaks being 3--4 times that for afterglows with 
achromatic breaks. This would suggest that, when the scattered emission is dominant at
X-rays, it is, on average, a few times brighter than that of the forward shock. For the
same set of X-ray afterglows with breaks, we find that the 7 chromatic breaks are preceded
by steeper decays ($\overline{\alpha_x} = 0.6\pm 0.2$) than the 7 achromatic breaks
($\overline{\alpha_x} = 0.1\pm 0.6$), that the achromatic breaks are stronger 
($<\delta \alpha_x^{(chr)}> = 0.7\pm 0.2$, $<\delta \alpha_x^{(achr)}> = 1.1\pm 0.7$), 
but that both types of light-curve breaks occur at about the same time (3-30 ks), 
which is somewhat surprising if they really have different origins.

\end{document}